\documentclass[prl,aps,floats,amsfonts,twocolumn,showpacs]{revtex4}

\usepackage{graphicx}

\newcommand{\beq}{\begin{equation}}
\newcommand{\eeq}{\end{equation}}
\newcommand{\beqa}{\begin{eqnarray}}
\newcommand{\eeqa}{\end{eqnarray}}

\def\half{\frac{1}{2}}
\def\<{\langle}
\def\>{\rangle}

\def\opone{\leavevmode\hbox{\small1\kern-3.8pt\normalsize1}}

\def\l{\vec\lambda}

\begin{document}

\title{Quantum entanglement can be simulated without communication}
\author{N. J. Cerf}
\affiliation{Quantum Information and Communication, Ecole
Polytechnique, CP 165/59, Universit\'e Libre de Bruxelles,
Avenue F. D. Roosevelt 50, 1050 Bruxelles, Belgium}
\author{N. Gisin}
\affiliation{GAP-Optique, University of Geneva,
20 rue de l'Ecole-de-M\'edecine, CH-1211, Switzerland}
\author{S. Massar}
\affiliation{Service de Physique Th\'eorique, CP 225, Universit\'e
Libre de Bruxelles, Boulevard du Triomphe, 1050 Bruxelles, Belgium}
\affiliation{Quantum Information and Communication, Ecole
Polytechnique, CP 165/59, Universit\'e Libre de Bruxelles,
Avenue F. D. Roosevelt 50, 1050 Bruxelles, Belgium}
\author{S. Popescu}
\affiliation{H. H. Wills Physics Laboratory, University of Bristol, Tyndall Avenue, Bristol BS8 1TL, U.K.}
\affiliation{Hewlett-Packard Laboratories, Stoke Gifford, Bristol BS12 6QZ, U.K.}

\begin{abstract}
It has recently been shown that all causal correlations
between two parties which output each one bit, $a$ and $b$, 
when receiving each one bit, $x$ and $y$,
can be expressed as convex combinations 
of local correlations (i.e., correlations that can be simulated 
with local random variables) and non-local correlations of the form
$a+b=x\cdot y \text{~mod~} 2$. We show that a single instance of 
the latter elementary non-local correlation 
suffices to simulate exactly all possible projective
measurements that can be performed on the singlet state of two
qubits, with no communication needed at all. This elementary non-local 
correlation thus defines some unit of non-locality, 
which we call a {\em nl-bit}.
\end{abstract}

\pacs{~}

\maketitle

The importance of quantum entanglement is by now widely
appreciated \cite{Terhal03}. Historically, entanglement has first
been viewed mainly as a source of paradoxes, most noticeably the 
Einstein-Podolsky-Rosen (EPR) paradox, 
which is at the origin of the concept of quantum non-locality
\cite{BellSpeakable}. Today, however, entanglement is rather viewed
as the resource that makes quantum information science so
successful \cite{LoSpillerPopescu98,physQI,NielsenChuang00}.
Indeed, based on entanglement, various informational tasks appear feasible,
although they would be impossible using only classical physics.

Following this new trend in quantum information science, a growing
community of physicists and computer scientists has started to investigate
the resource ``entanglement''. Questions like how 
to manipulate this resource, e.g., how to concentrate or dilute it
\cite{BennettPopescu96}, or how to transform
it into secret bits \cite{Curty04,AcinGisin03}, were addressed. 
Also, a unit of entanglement has been identified and named {\em e-bit};
it consists of a pair of maximally entangled qubits, e.g., a singlet
as used in Bohm's version of the EPR paradox. 
A few years ago, connections with communication complexity started to be
studied \cite{Brassard01}, with questions like how much classical
communication is required to simulate an e-bit?

\begin{figure}[h]
\includegraphics[width=7cm]{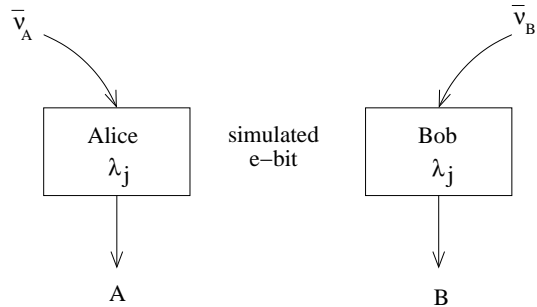}
\caption{{Principle of e-bit simulation. The statistics of the
output bits A and B should coincide with that predicted by
quantum physics for the measurements defined by $\vec\nu_A$ and
$\vec\nu_B$. The $\lambda_j$ denote random data that Alice and Bob
can share beforehand, when they jointly agreed on a strategy. The
inputs $\vec\nu_A$ and $\vec\nu_B$ are given to Alice and to Bob,
respectively, after they separated. Note that each party is oblivious
of the other party's input.}} \label{simul}
\end{figure}

Simulating an e-bit means the following. Two parties, Alice and
Bob, receive each a normalized vector $\vec\nu_A$ and $\vec\nu_B$ 
that characterizes
their measurement on the Poincar\'e sphere, and each has to
output a bit, $A$ and $B$ \cite{bit}, see Fig. \ref{simul}. 
The statistics of the output bits should exactly reproduce 
the quantum predictions for all values of $\vec\nu_A$ and $\vec\nu_B$
if Alice and Bob were actually 
sharing a singlet state $(|01\rangle-|10\rangle)/\sqrt{2}$.
For instance, if the vectors are opposite, $\vec\nu_A=-\vec\nu_B$, the
output bits should always be equal, $A=B$. From Bell inequality, we
know that it is impossible to simulate a singlet without any
communication. This is so even if one assumes that both
parties share local hidden variables, or in modern terminology,
local randomness (that is, they share a non-finite list of random bits
$\lambda_j$). Of course, if an unlimited amount of
communication is allowed, then Alice could simply send her measurement
setting $\vec\nu_A$ to Bob with arbitrary precision, so the simulation 
of a singlet would become straightforward. 
But whether such an unlimited amount
of communication is necessary was unknown. First answers along 
this direction were given by A. Tapp, R. Cleve, and G. Brassard \cite{TappBrassard99} in Montreal, and by M. Steiner \cite{Steiner00} 
from the NSA. The Canadian group showed that, quite surprisingly, 
8 bits of communication suffice for a perfect (analytic) simulation
of the quantum predictions. Steiner, followed by \cite{GisinGisin99}, 
showed that if one allows the number of bits to vary from one instance 
to another, then 2 bits suffice on average. It was also shown that,
with block coding, the number of communicated bits can be reduced
to 1.19 bits on average \cite{cerfetal}. A few years later, 
B. Toner and D. Bacon \cite{TonerBacon03} improved on these
results and showed that actually a single bit of communication
suffices for perfect simulation of a singlet. At this point,
the situation was the following: one bit of communication allows 
one to simulate a singlet, and one singlet provides one secret bit.

\begin{figure}[h]
\includegraphics[width=7cm]{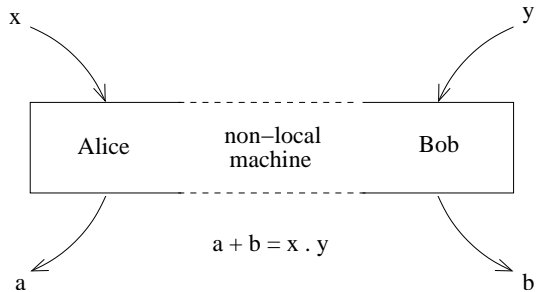}
\caption{{Scheme of the PR non-local machine, where $x,y$ and $a,b$
denote the input and output bits, respectively.}} \label{scheme}
\end{figure}

Independently of the above story, S. Popescu and D. Rohrlich raised
the following question: can there be stronger correlations 
than the quantum mechanical ones that remain causal (i.e., that
do not allow signaling) \cite{PopescuRohrlich97}? 
Recall that the quantum correlations violate the Bell inequality, 
but do not allow any faster than light signaling.
Popescu and Rohrlich answered by presenting an hypothetical machine
that does not allow signaling, yet violates the Clauser-Horne-Shimony-Holt (CHSH) \cite{CHSH} inequality more than quantum mechanics.
They concluded by asking why Nature is non-local, but not maximally non-local,
where the maximum would only be limited by the no-signaling constraint?

In this Letter, we push this investigation even further by showing that,
actually, quantum entanglement {\em can} be perfectly simulated
by using one instance of this non-local PR machine and {\em no}
communication at all! Since, as we will show, one instance of the PR machine
is a weaker resource than one bit of communication, one is tempted to
conclude that Nature may use something like these non-local machines 
if she is sparing with resources.

\paragraph{Non-local PR machine.}
The non-local PR machine works as follows, see Fig. \ref{scheme}.
It admits two input bits $x$ and $y$, and yields two output bits $a$ and
$b$. The bits $x$ and $a$ are in Alice's hands, while $y$ and $b$
are on Bob's side. The machine is such that $a$ and $b$ are correlated 
according to the simple relation (equality modulo 2): 
\beq a+b=x\cdot y
\label{NLrelation}
\eeq
Except for this relation, $a$ and $b$ are unbiased random bits. 
For example, if $x=y=0$, then the machine's outputs are
random but identical: $a=b=0$ or $a=b=1$ with equal probabilities
$\half$. This implies that the PR machine cannot be used to signal: since
the output $a$ ($b$) is locally random, its value cannot convey any
information about the input $y$ ($x$) of the other party.
This machine is constructed in such a way that the CHSH inequality 
is violated by the algebraic maximum value of 4, while
quantum correlation achieve at most $2\sqrt{2}$ \cite{Cirelson80}. 
(Remember that with shared randomness only, the maximum allowed value 
in a local theory is 2.) To see this, let us change the bit values
0 and 1 to the values $\pm 1$ traditionally used in Bell
inequalities. Defines $a'=1-2a$ and $b'=1-2b$ and note that
\begin{equation}
a'\cdot b'=\left\{
\begin{tabular}{ll}
1 & \text{if $a+b=0 \text{~mod~}2$,}\\
-1 & \text{if $a+b=1 \text{~mod~}2$.}
\end{tabular}
\right.
\end{equation}
Denoting by $E$ the expectation value, one has for the CHSH
inequality: $E(a'\cdot b'|x=0,y=0)+E(a'\cdot b'|x=0,y=1)+E(a'\cdot
b'|x=1,y=0)-E(a'\cdot b'|x=1,y=1)=4>2$. The violation of CHSH
inequality implies that this PR machine is non-local (even more
than quantum physics), so that it cannot be simulated with
local variables. Yet, it is causal, like quantum mechanics.

Let us emphasize that the PR machine (\ref{NLrelation}) is not
an arbitrary construction. It is, up to elementary 
symmetries like bit flips, the unique binary causal 
{\em maximally} non-local machine. Indeed, it can be shown
that all binary causal correlations can be expressed as convex combinations
of local machines (i.e., those which can be simulated with local random variables) and maximally non-local PR machines \cite{Barrett04}. 
The PR machines also have the surprising property that, given an unlimited
supply of them, any communication complexity problem can be
solved with a single bit of communication \cite{vanDam00}. 

Finally, note that it is straightforward to simulate a PR machine with
shared randomness (i.e., local hidden variables) augmented by one bit of
communication: the hidden variable $\lambda$ should then be a random unbiased
bit, $a=\lambda$, and $x$ should be communicated by Alice to Bob who
should output $b=x\cdot y+\lambda \text{~mod~}2$. 
But the converse is false: a PR machine
cannot be used to communicate since it is causal. Therefore, as already
mentioned, the PR machine is a strictly weaker resource than a bit of communication, that is
\beq
1 \text{~nl-bit} \prec 1 \text{~bit~(supraluminal~communication)}
\label{prec}
\eeq
where we have denoted as {\em nl-bit} the unit of non-local correlations
effected by the PR machine.

\paragraph{Simulation of a singlet with a non-local PR machine.}
We now show that any projective measurements on a singlet
can be perfectly simulated using a single instance 
of this non-local PR machine, with no communication being necessary.
As a consequence of (\ref{prec}),
this is a stronger result than the simulation of a singlet 
with one communicated bit \cite{TonerBacon03}.
Consider that Alice and Bob share a non-local PR machine
as well as shared randomness in the form of pairs of normalized vectors $\l_1$
and $\l_2$, randomly and independently distributed over the entire
Poincar\'e sphere. Denote $\vec\nu_A$ and $\vec\nu_B$ the vectors
that determine Alice and Bob measurements, respectively.

The model goes as follows. Alice inputs
\begin{equation}
x=sg(\vec \nu_A\cdot\l_1)+sg(\vec\nu_A\cdot\l_2)
\end{equation}
into the machine, where
\begin{equation}
sg(x)=\left\{
\begin{tabular}{ll}
1 & \text{if $x\ge 0$,}\\
0 & \text{if $x<0$.}
\end{tabular}
\right.
\end{equation}
(Here and now on, all equalities involving bits are taken modulo 2.)
She then receives the bit $a$ out of the machine, and
outputs 
\begin{equation}
A=a+sg(\vec \nu_A\cdot\l_1)
\end{equation}
as the {\it simulated} measurement outcome. Similarly, Bob inputs
\begin{equation}
y=sg(\vec\nu_B\cdot\l_+)+sg(\vec\nu_B\cdot\l_-)
\end{equation}
into the machine, where $\l_\pm=\l_1\pm\l_2$, receives $b$ out of the
machine, and then outputs
\begin{equation}
B=b+sg(\vec\nu_B\cdot\l_+)+1.
\end{equation}

Note that since the machine's outputs $a$ and $b$ are random unbiased bits,
the simulated measurement outcomes $A$ and $B$ are equally random,
exactly as for real measurements on a singlet. But
the outputs $a$ and $b$ are correlated according to relation
(\ref{NLrelation}), hence $A$ and $B$ are also correlated. The
surprising and interesting result is that this correlation is
precisely the one predicted by quantum mechanics for the singlet
state:\\
{\bf Theorem}:\hspace{1cm} \beq
E(A+B|\vec\nu_A,\vec\nu_B) = 
\frac{1+\vec\nu_A\cdot\vec\nu_B}{2}\label{singletCorrel}\eeq
 {\em
Proof}: First, compute
\begin{eqnarray}
A+B&=&a+b+sg(\vec \nu_A.\vec\lambda_1)+sg(\vec \nu_B.\vec\lambda_+)+1
 \nonumber\\
&=&x\cdot y+sg(\vec \nu_A.\vec\lambda_1)+sg(\vec\nu_B.\vec\lambda_+)+1
 \nonumber \\
&=&z+sg(\vec \nu_A.\vec\lambda_1)+sg(\vec\nu_B.\vec\lambda_+)+1
\label{TonerBacon}
\end{eqnarray}
where
\begin{equation}
z=[sg(\vec \nu_A.\vec\lambda_1)+sg(\vec \nu_A.\vec\lambda_2)]
[sg(\vec \nu_B.\vec\lambda_+) + sg(\vec \nu_B.\vec\lambda_-) ]
\end{equation}
Next, note that (\ref{TonerBacon}) corresponds precisely to the
1-bit communication model \cite{TonerBacon03}. Indeed, in
this model, Alice outputs $A= sg(\vec \nu_A.\vec\lambda_1)$,
communicates the bit $c=sg(\vec \nu_A.\vec\lambda_1)+sg(\vec
\nu_A.\vec\lambda_2)$ to Bob who outputs $B=(1-c) \;  sg(\vec\nu_B.\vec\lambda_+) + c \; sg(\vec \nu_B.\vec\lambda_-)+1$. 
The latter can be re-expressed as $B = z+sg(\vec \nu_B.\vec\lambda_+)+1$. 
Thus, $A+B=z+sg(\vec \nu_A.\vec\lambda_1)+sg(\vec\nu_B.\vec\lambda_+)+1$. Finally, since the expressions for $A+B$ in our model
and the 1-bit communication model are identical
and since the latter model satisfies (\ref{singletCorrel}),
so does our model \cite{coding}.$\Box$

\paragraph{Analogue of entanglement monogamy:
the non-local PR machine cannot be shared.}
Given the analogy between the entanglement contained in a singlet
(1 e-bit) and the non-local but causal correlations produced by the PR
machine (1 nl-bit), it is tempting to investigate how deep this analogy
can be pushed. One of the key features of entanglement is its {\it monogamy}
\cite{Terhal03}. By this one means that if a quantum system A is
strongly entangled with another system B, then A cannot simultaneously
share much entanglement with any third system C.
This property is for example at the basis of the quantum no-cloning theorem
\cite{Dieks82}, the monogamy of CHSH inequalities \cite{ScaraniGisin},
or the security of quantum cryptography \cite{ShorPreskill00}. 
We shall see that this same property holds for causal non-local machines.

\begin{figure}[h]
\includegraphics[width=7cm]{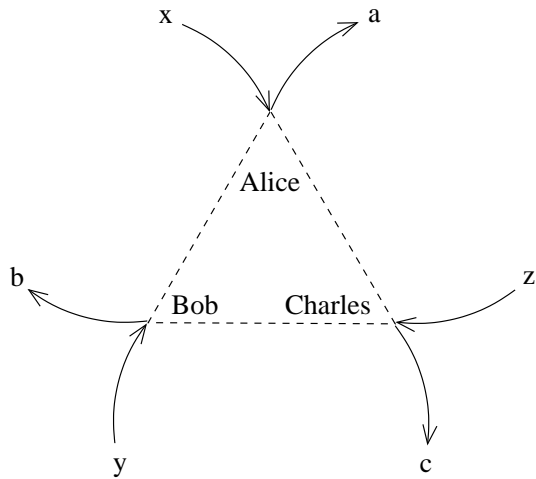}
\caption{{Scheme of a 3-party nonlocal machine.}}
\label{scheme3party}
\end{figure}

First, let us summarize the argument of \cite{Dieks82} underlying
the monogamy of entanglement in order to emphasize the analogy with our result
for causal non-local machines. Consider Alice and Bob share a pair of 
maximally-entangled qubits. Assume that Bob is able to perfectly duplicate
his qubit and make two clones (one that he keeps for himself, and the other one
that he passes to Charles), so that Alice's qubit is now part of a singlet 
state both with Bob and Charles. Then, by measuring her qubit in either the 
computational basis or the dual basis, Alice would prepare the 2-qubit system
shared by Bob and Charles in two different mixtures, which would
allow instantaneous signaling between Alice and Bob/Charles. Hence,
perfect cloning is impossible, and entanglement must be monogamous.
Now, coming back to the monogamy of causal non-local machines,
assume Alice holds the two halves of two PR machines, one
shared with Bob, the other one shared with Charles 
(see Fig.~\ref{scheme3party}). Denote by $z$ and
$c$ Charles' input and output bits. One has
\begin{eqnarray}
a+b&=&x\, y   \nonumber \\ a+c&=&x\, z
\end{eqnarray}
Therefore, we have $b+c=x(y+z)$. Assume now that Bob and Charles 
sit next to each other, at a long distance from Alice. Then if Bob
enters $y=0$ and Charles enters $z=1$ in their respective
machines, we have $b+c=x$. This means that, by checking whether
their outputs are equal or not, Bob and Charles can know
instantaneously whether Alice entered $x=0$ or $x=1$ into the
machine.  Such a tripartite PR machine would then provide
instantaneous signaling between Alice and Bob/Charles. Hence,
it cannot exist, and causal non-local machines must be monogamous.

\paragraph{Conclusion.}
Quantum non-locality is one of the most important and amazing
discoveries of the 20th century physics. It took a long time to be
appreciated, and actually it is still believed to contain deep
mysteries. However, today, thanks to quantum information science,
entanglement has become better studied and understood. Probably
its most remarkable manifestation is quantum teleportation
\cite{teleportation}, a protocol that allows one to teleport all
the characteristics of an object embedded in some energy and
matter localized ``here'' to another piece of energy and matter
located at a distance. In this Letter, we contributed to
``disentangle'' the non-locality inherent to quantum mechanics
into its elementary constituent, a {\em unit of non-locality} or {\em nl-bit}. 
Surprisingly, the quantum non-locality of a singlet boils down 
to a rather simple machine, encapsulated by relation (\ref{NLrelation}),
which is inspired by the CHSH inequality. We showed that one instance of
this non-local machine is sufficient to perfectly simulate a singlet.
Since this machine defines a resource that is strictly weaker
than any communication while it is sufficient to simulate a singlet, 
we have in short
\beq
1 \text{~e-bit~(simulation of)}  \prec 1  \text{~nl-bit} \prec 1 \text{~bit}
\label{prec2}
\eeq
Thus, assuming that Nature is sparing with resources, 
one is tempted to conclude that the non-local correlations that she exhibits
originate from these kinds of machines.


\begin{thebibliography}{99}
\bibitem{bit} We adopt the computer science terminology: bit
values 0 and 1 instead of the spin values $\pm\half$ or the values
$\pm1$ often used in the context of Bell inequalities.
\bibitem{coding} Note that, conditionally on $\vec\lambda_1$ and
$\vec\lambda_2$, the bit $c$ that is communicated in the model of
\cite{TonerBacon03} is not equiprobable. Therefore, by using block coding,
less than one bit must be communicated on average to simulate a singlet.
In our model, this translates into the fact that we do not fully use
the non-locality of the PR machine to simulate a singlet.
\end{thebibliography}
\end{document}